# SELF-COLLAPSE AND SLIDING OF NANOTUBES IN A BUNDLE

Nicola Maria Pugno
*Politecnico di Torino, Italy*

**Summary**

We have discovered that the influence of the surrounding nanotubes in a bundle is similar to that of a liquid having surface tension equal to the surface energy of the nanotubes. This surprising behaviour is confirmed by the calculation of the self-collapse diameters of nanotubes in a bundle. Other systems, such as peapods, fullerites, are similarly treated, including the effect of the presence of a solvent. Finally, we have evaluated the strength and toughness of the nanotube bundle, with or without collapsed nanotubes, assuming a sliding failure.

**1. On the polygonization, collapse, self-collapse and "dog-bone" configurations of an isolated nanotube or of nanotubes in a bundle**

*Polygonization*

Due to surface energy (mainly van der Waals attraction) and/or external pressure the nanotubes in a bundle tend to polygonize, from the circular towards the hexagonal shape, Figure 1.

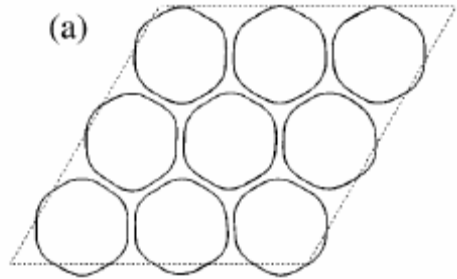

Figure 1: Polygonization of nanotubes in a bundle, calculated according to atomist simulations (from Elliot et al., Physical Review Letters, 2004, 92, 095501).

In general, a blunt hexagon is expected for a nanotube with a small number of walls, e.g., single, double or triple walled nanotubes. Let us indicate the radius of the blunt notches with $r$ and the length of the rectilinear sides with $a$. Denoting with $R$ the nanotube radius, the inextensibility condition implies $2\pi R = 6a + 2\pi r$, from which we deduce:

$$r(a) = R - \frac{3}{\pi}a \qquad (1)$$

The nanotube cross-sectional area is:

$$A(a) = 6ar + \pi r^2 + \frac{3}{2}\sqrt{3}a^2 = 6a\left(R - \frac{3}{\pi}a\right) + \pi\left(R - \frac{3}{\pi}a\right)^2 + \frac{3}{2}\sqrt{3}a^2 \qquad (2)$$

and consequently $a(A) = \sqrt{\dfrac{\pi R^2 - A}{c/2}}$, with $c = \dfrac{18}{\pi} - 3\sqrt{3} \approx 0.54$. Posing $A_0 = \pi R^2$ and $\Delta A = A_0 - A$, we can introduce $0 \leq a^* \equiv \dfrac{a}{R} \leq \dfrac{\pi}{3}$:

$$a^* = \frac{a}{R} = \sqrt{\frac{2\pi}{c}} \sqrt{\frac{\Delta A}{A_0}} \qquad (3)$$

Note that $a^*$ is proportional to the relative contact length as well as to the relative area/volume variation.

The equilibrium of the system can be calculated by minimizing its free energy. Indicating with $\Phi$ the elastic energy, with $\gamma$ the surface energy and with $p$ the external (relative) pressure, the energy minimization implies:

$$\mathrm{d}\Phi + pL\mathrm{d}A - 6\gamma L \mathrm{d}a = 0 \qquad (4)$$

where $L$ is the nanotube length.

According to elasticity, the strain energy stored per unit surface is $\dfrac{\mathrm{d}\Phi}{Lr\mathrm{d}\vartheta} = \dfrac{N^\alpha D}{2r^2}$ ($\vartheta$ is the angle describing the surface at the radius $r$), where:

$$D = \frac{Et^3}{12\sqrt{1-v^2}} \qquad (5)$$

is the nanotube bending rigidity, $N$ is the number of walls and $1 \leq \alpha \leq 3$: assuming perfect bonding between the walls would correspond to $\alpha = 3$, whereas for independent walls $\alpha = 1$ (however note that in the equations appears always the group $N^\alpha D$, that is the total bending stiffness); $E$ is the Young' modulus of graphene and $t \approx 0.34$nm is the conventional wall thickness.

Accordingly, $\Phi = \dfrac{\pi D L}{r}$ and $\dfrac{\mathrm{d}\Phi}{\mathrm{d}a} = \dfrac{3 N^\alpha D L}{r^2}$. In addition, $\dfrac{\mathrm{d}A}{\mathrm{d}a} = -ca$. Thus, the free energy minimization yields:

$$p(a^*) = \frac{3 N^\alpha D}{c a^* \left(1 - \dfrac{3}{\pi} a^*\right)^2 R^3} - \frac{6\gamma}{c a^* R} \qquad (6)$$

Under zero pressure the equilibrium is reached in the following configuration:

$$a_0^* = a^*(p=0) = \frac{\pi}{3}\left(1 - \frac{1}{R}\sqrt{\frac{N^\alpha D}{2\gamma}}\right) \qquad (7)$$

showing that for radii:

$$R \leq R_0^{(N)} = \sqrt{\frac{N^\alpha D}{2\gamma}} \qquad (8)$$

the contact length is physically zero (mathematically it is negative) and thus the surface energy is not capable of producing even an infinitesimal polygonization in so small nanotubes. This peculiarity, of zero contact length for small radii, is also observed during the adhesion of single walled nanotubes over a flat substrate (N. Pugno, *An analogy between the adhesion of liquid drops and single walled nanotubes,* SCRIPTA MATERIALIA (2008) 58, 73-75). Taking $D = 0.11\text{nN} \cdot \text{nm}$ (bending stiffness of graphene $D \approx 0.9 - 0.24 \text{nN} \cdot \text{nm}$) and $\gamma = 0.18 \text{N/m}$ (surface energy of graphene $\gamma \approx 0.16 - 0.20 \text{N/m}$), we find $2R_0^{(1)} \approx 1.1\text{nm}$. Assuming an intermediate coupling between the walls, i.e. $\alpha \approx 2$, the critical diameters for double and triple walled nanotubes are $2R_0^{(2)} \approx 2.2\text{nm}$ and $2R_0^{(3)} \approx 3.3\text{nm}$.

For larger nanotubes, the adhesion energy induces a polygonization, as described by eq. (7). The action of an external pressure further increases the polygonization, according to the state equation (6), see Figure 2.

Note that eqs. (1) and (7) imply that under zero pressure the blunt radius $r$ assumes the constant value $R_0^{(N)}$, as defined in eq. (8).

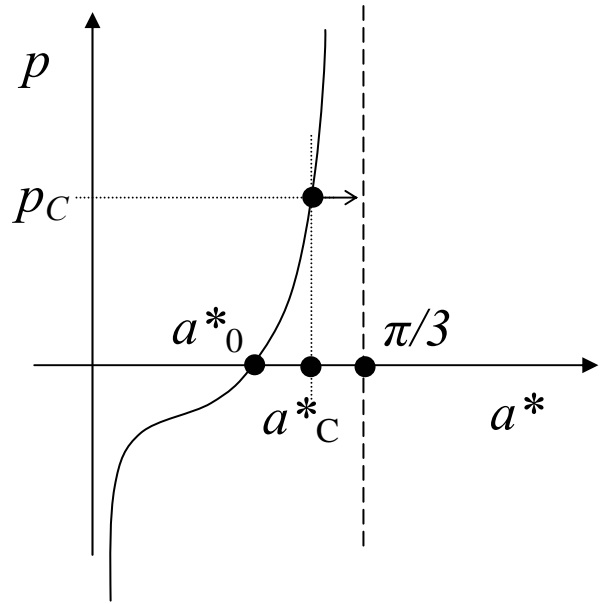

Figure 2: State equation (6) for the polygonization of nanotubes in a bundle
(an inflection point appears at a negative pressure, but the curve has everywhere a positive slope, thus the process is stable).

*Collapse*

Eq. (6) correctly predicts that to reach a full polygonization the pressure must tend to infinity, as the elastic energy stored in sharp notches, namely $p \to \infty$ for $a_0^* \to \dfrac{\pi}{3}$; practically, a different mechanism, that is the well-know elastic instability, Figure 3, will take place at a finite value of the applied pressure.

We treat the large nanotube bundle as a liquid-like material with surface tension $\gamma_t = \gamma$, as imposed by the energy equivalence (the surface tension has the thermodynamic significance of work spent to create the unit surface, as the surface energy), thus deducing a pressure $\gamma/R$ acting on a single nanotube of radius $R$ within a bundle, as evinced by the Laplace's equation. The critical pressure can be accordingly derived as:

$$p_C = \frac{3N^\alpha D}{R^3} - \frac{\gamma}{R} \qquad (9)$$

The first term in eq. (9) is that governing the buckling of a perfectly elastic cylindrical long thin shell (of thickness $Nt$), whereas the second term is the pressure imposed by the surrounding nanotubes.

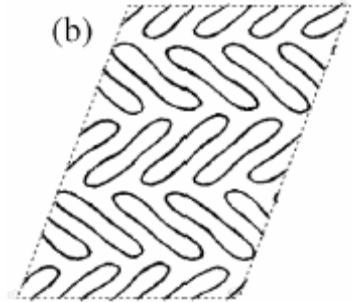

Figure 3: Collapse of nanotubes in a bundle, calculated according to atomist simulations (from Elliot et al., Physical Review Letters, 2004, 92, 095501).

Treating the single walled nanotube data by Elliot et al. (Physical Review Letters, 2004, 92, 095501), excluding the two smallest nanotubes for which the buckling pressure was not accurately determined, a relevant agreement with eq. (9) is observed (coefficient of correlation $R^2$=0.97), fitting a plausible value of $D_{fit} \approx 0.2 \text{nN} \cdot \text{nm}$, see Figure 4.

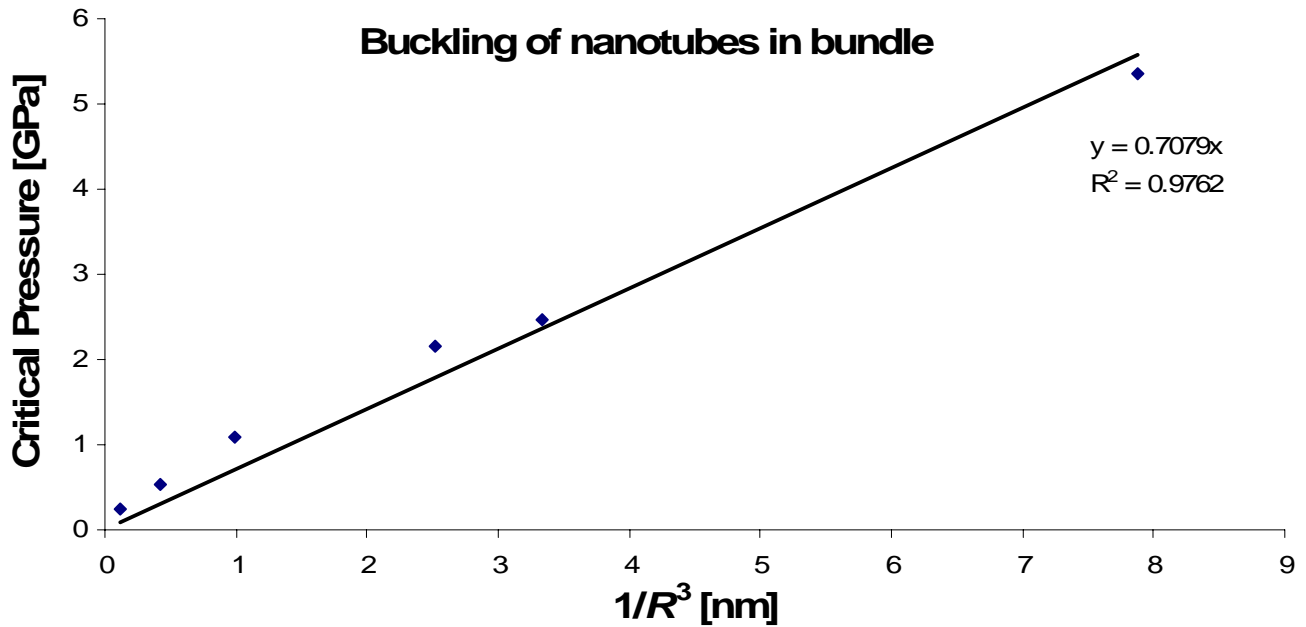

Figure 4: Collapse of nanotubes in a bundle, calculated according to atomist simulations (from Elliot et al., Physical Review Letters, 2004, 92, 095501); total critical pressure $p_C + \frac{\gamma}{R}$ versus $\frac{1}{R^3}$, the slope is thus $3D$.

Other atomist simulation results (from Elliot) are compared with theoretical predictions in Table 1, for both single and double walled nanotubes.

| Nanotube | Diameter [nm] | Collapse pressure [GPa] Atomistic Simulations | Collapse pressure [GPa] Theory |
|---|---|---|---|
| (11,11) | 1.49 | 1.91 | 1.90 |
| (13,9), | 1.50 | 2.03 | 1.87 |
| (19,0) | 1.49 | 1.99 | 1.92 |
| (16,5) | 1.49 | 1.98 | 1.90 |
| (5,5) | 0.6 | - | 22.51 |
| (7,7) | 0.95 | 5.50 | 8.00 |
| (10,10) | 1.36 | 2.49 | 2.59 |
| (15,15) | 2.03 | 0.95 | 0.66 |
| {(15,15),(10,10)} | 1.70 | 3.45 | 2.48 |
| {10,10),(5,5)} | 1.02 | 9.47 | 12.90 |
| {(11,11),(7,7)} | 1.22 | 6.80-7.20 | 7.26 |
| {(15,15),(7,7)} | 1.49 | 1.00-4.00 | 3.80 |

Table 1: Collapse pressures for single and double walled nanotubes, calculated according to atomistic simulations and theory (eq. (3) with $D = 0.3\,\text{nN}\cdot\text{nm}$, $\alpha = 2$ and $\gamma = 0.2\,\text{N/m}$; the slightly larger value of $D$ may suggest a slightly larger numerical coefficient than the factor of 3 posed by elasticity; an intermediate walls interaction is deduced) .

*Self-collapse*

From eq. (9) we derive the following condition for the self-collapse, i.e. collapse under zero pressure, of a nanotube in a bundle:

$$R \geq R_C^{(N)} = \sqrt{\frac{3N^\alpha D}{\gamma}} = \sqrt{6} R_0^{(N)} \qquad (10)$$

Eq. (10) assumes for the self-collapse a vanishing difference in pressure between the inner and outer environments of the nanotube; however, if vacuum is internally present, the self-collapse pressure must be considered not zero but the atmospheric pressure $p_A \approx 0.1\,\text{MPa}$. However this value is small and does not significantly affect the prediction of eq. (10). In fact, new self-collapse radius can be calculated according to eq. (9) with $p_C = p_A$, solving the corresponding third-order polynomial equation. However, a correction with respect to the previously evaluated self-collapse radius $R_C^{(N)}$ can be considered inserting into eq. (10) $R \to R(1+\varepsilon)$, neglecting the powers of $\varepsilon$ higher than one and noting that $R_C^{(N)}$ is the solution of the equation for $p_C = 0$; we accordingly find $\varepsilon = -\frac{c a_C^* R_C^{(N)} p_A}{6\gamma}$; this number is of the order of $\varepsilon \approx -R_C p_A/\gamma \approx -10^{-4}$ and confirms the hypothesis.

Taking $D = 0.11\,\text{nN}\cdot\text{nm}$ and $\gamma = 0.18\,\text{N/m}$ we find $2R_C^{(1)} \approx 2.7\,\text{nm}$. Considering an intermediate coupling between the walls ($\alpha \approx 2$), the critical diameters for double and triple walled nanotubes are $2R_C^{(2)} \approx 5.4\,\text{nm}$ and $2R_C^{(3)} \approx 8.1\,\text{nm}$.

In a recent paper by the Prof. Windle's group (Motta et al., Advanced Materials, 2007, 19, 3721), 17 experimental observations on the self-collapse of nanotubes in a bundle have been reported, see Figure 5 and Table 2. A number of 5 single walled nanotubes with diameters in the range 4.6-5.7nm were all observed as collapsed; moreover, while the 3 double walled nanotubes observed with internal diameters in the range 4.2-4.7nm (the effective diameters are larger by a factor of ~0.34/2nm) had not collapsed, the observed 8 double walled nanotubes with internal diameters in the range 6.2-8.4nm had collapsed. Finally, a triple walled nanotube of 14nm internal diameter (the effective diameter is ~14.34m) was observed as collapsed too. All these 17 observations are in agreement with our theoretical predictions of eq. (10).

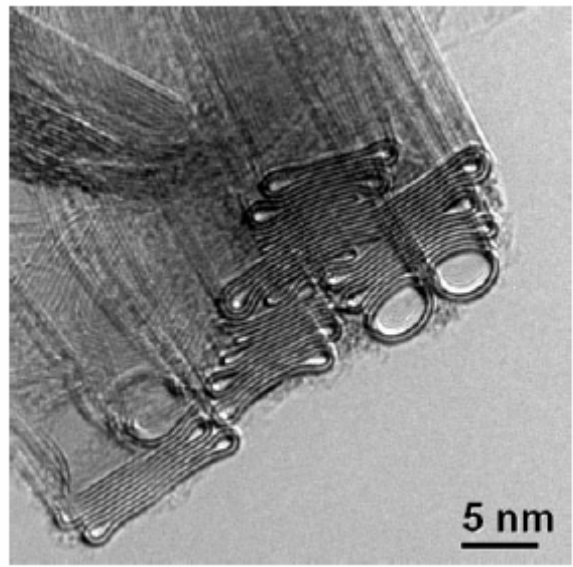

Figure 4: Self-collapsed nanotubes in a bundle
(from Motta et al., Advanced Materials, 2007, 19, 3721).

| Nanotube number | Number $N$ of walls | Diameter of the internal wall [nm] | Collapsed (Y/N) Exp. & Theo. |
|---|---|---|---|
| 1 | 1 | 4.6 | Y |
| 2 | 1 | 4.7 | Y |
| 3 | 1 | 4.8 | Y |
| 4 | 1 | 5.2 | Y |
| 5 | 1 | 5.7 | Y |
| 6 | 2 | 4.2 | N |
| 7 | 2 | 4.6 | N |
| 8 | 2 | 4.7 | N |
| 9 | 2 | 6.2 | Y |
| 10 | 2 | 6.5 | Y |
| 11 | 2 | 6.8 | Y |
| 12 | 2 | 6.8 | Y |
| 13 | 2 | 7.9 | Y |
| 14 | 2 | 8.3 | Y |
| 15 | 2 | 8.3 | Y |
| 16 | 2 | 8.4 | Y |
| 17 | 3 | 14.0 | Y |

Table 2: Self-collapse of nanotubes in a bundle: theory exactly fits the experimental observations (data from Motta et al., Advanced Materials, 2007, 19, 3721).

*"Dog-bone" configuration*

The collapsed nanotubes assume a characteristic "dog-bone" cross-sectional shape, since the radius of curvature cannot be infinitely small, see Figures 4 and 5.

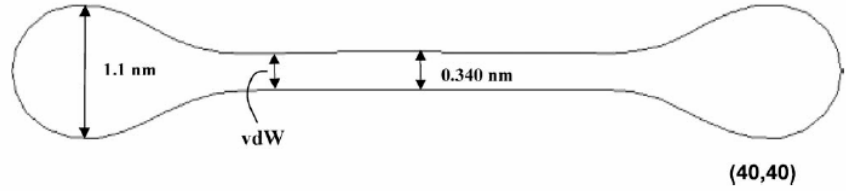

Figure 5: "Dog-bone" configuration, calculated according to finite element simulations (from Pantano et al., Journal of the Mechanics and Physics of Solids, 2004, 52, 789).

In order to derive the equilibrium of the dog-bone configuration, the free energy minimization can again be considered. Let us indicate with $r$ the radius of the two terminal lobes and with $a$ the length of the two rectilinear sides in mutual contact. Denoting with $R$ the nanotube radius, the inextensibility condition implies $2\pi R = 2a + 4\pi r$, from which we deduce:

$$r = \frac{R}{2} - \frac{a}{2\pi} \tag{11}$$

The nanotube cross-sectional area is:

$$A = 2\pi r^2 \tag{12}$$

The energy minimization implies ($\Phi = \frac{2\pi N^\alpha DL}{r}$):

$$\frac{d\Phi}{dr} + pL\frac{dA}{dr} - 2\gamma L\frac{da}{dr} = 0 \tag{13}$$

corresponding to the following equilibrium:

$$p(r) = \frac{N^\alpha D}{2r^3} - \frac{\gamma}{r} \tag{14}$$

For zero surface energy the equilibrium pressure is positive (inward), whereas complimentary for zero bending stiffness it is negative (outward). Under zero pressure the equilibrium is reached for:

$$r_0 = \sqrt{\frac{N^\alpha D}{2\gamma}} = R_0^{(N)} \tag{15}$$

In such a case, we predict an equilibrium diameter for a single walled nanotube of $2r_0 \approx 1.1 \text{nm}$, in perfect agreement with previous calculations (see Figure 4).

Posing $a=0$ in eq. (11) and comparing with eq. (14), we deduce the critical pressure corresponding to the dog-bone "opening" (since the process is stable, see in the following):

$$p_O = \frac{4N^\alpha D}{R^3} - \frac{2\gamma}{R} \qquad (16)$$

Moreover, posing $p_O = 0$ in eq. (16) suggests that for nanotube radii:

$$R \leq \sqrt{\frac{2N^\alpha D}{\gamma}} = 2R_0^{(N)} \qquad (17)$$

the "dog-bone" configuration cannot be self-maintained (for single nanotubes such a diameter is $2R_O \approx 2.2 \text{nm}$).

Note that eq. (14) presents an inflection point with zero slope at $r = \sqrt{3}r_0$, suggesting that at such a point a rapid change in the configuration will take place at an "anti-buckling" pressure $p(r=\sqrt{3}r_0) = -\frac{2\gamma}{3\sqrt{3}r_0} = -\frac{4}{3\sqrt{6N^\alpha}}\sqrt{\frac{\gamma^3}{D}}$.

If the nanotube is assumed to be in contact with other adjacent nanotubes along its two external sides of length $a$, the equations presented in this section are still valid with the substitution $\gamma \to 2\gamma$.

## 2. Calculation of the strength and toughness of nanotube bundles under sliding failure mode

*Strength*

The fracture mechanics approach developed by Pugno in several papers (e.g. see for instance [1,2]) could be of interest to evaluate the strength of nanotube bundles assuming a sliding failure mode [3]. This hypothesis is complimentary to that of intrinsic nanotube fracture, already treated by the same author in numerous papers (e.g. see for instance [4-6]). Thus we assume the interactions between adjacent nanotubes as the weakest-links, i.e. that the fracture of the bundle is caused by nanotube sliding rather than by their intrinsic fracture.

Accordingly, the energy balance during a longitudinal delamination d$z$ under the applied force $F$, is:

$$d\Phi - Fdu - 2\gamma(P_C + P_{vdW})dz = 0 \qquad (1)$$

where $d\Phi$ and $du$ are the strain energy and elastic displacement variation due to the infinitesimal increment in the compliance caused by the delamination d$z$; $P_{vdW}$ describes the still existing van der Waals interaction for vanishing nominal contact nanotube perimeter $P_C = 6a$ (the shear force between two graphite single layers becomes zero for nominally negative contact area); $6a$ is the contact length due to polygonization of nanotubes in the bundle, caused by their surface energy $\gamma$.

Elasticity poses $\frac{d\Phi}{dz} = -\frac{F^2}{2ES}$, where $S$ is the cross-sectional surface area of the nanotube, whereas according to Clapeyron's theorem $Fdu = 2d\Phi$. Thus, the following simple expression for the bundle strength ($\sigma_C = F_C/S$, effective stress and cross-sectional surface area are here considered; $F_C$ is the force at fracture) is predicted:

$$\sigma_C^{(theo)} = 2\sqrt{E\gamma\frac{P}{S}} \qquad (2)$$

in which it appears the ratio between the effective perimeter ($P = P_C + P_{vdW}$) in contact and the cross-sectional surface area of the nanotubes.
Note that eq. (2) is basically the asymptotic limit for sufficiently long overlapping length; for overlapping length smaller than a critical value the strength increases by increasing the overlapping length, see [2,3]; for a single nanotube this overlapping length is of the order of 10 microns, whereas it is expected to be larger for nanotubes in bundles, e.g. of the order of several millimetres, as confirmed experimentally, see Vilatela's PhD Thesis). This critical length is [2,3]:

$$\ell_C \approx 6\sqrt{\frac{hES}{PG}} \qquad (3)$$

where $h$ and $G$ are the thickness and shear modulus of the interface. Eq. (3) suggests that increasing the size-scale $L \propto \sqrt{S} \propto P \propto h$ this critical length increases too, namely $\ell \propto L$, thus the strength increases by increasing the overlapping length in a wider range; however note that the achievable strength is reduced since, $\sigma_C^{(theo)} \propto \ell^{-1} \propto \sqrt{P/S}$ : *increasing the overlapping length ad infinitum is not a way to indefinitely increases the strength.*

The real strength could be significantly smaller, than that predicted by eq. (3), not only because $\ell < \ell_C$ but also as a consequence of the misalignment of the nanotubes with respect to the bundle axis. Assuming a non perfect alignment of the nanotubes in the bundle, described by a non zero angle $\beta$, the longitudinal force carried by the nanotubes will be $F/\cos\beta$, thus the equivalent Young' modulus of the bundle will be $E\cos^2\beta$, as can be evinced by the corresponding modification of the energy balance during delamination; accordingly:

$$\sigma_C = 2\cos\beta\sqrt{E\gamma\frac{P}{S}} \qquad (4)$$

The maximal achievable strength is predicted for collapsed perfectly aligned (sufficiently overlapped) nanotubes, i.e. $\frac{P}{S} \approx \frac{1}{Nt}$, $\beta = 0$:

$$\sigma_C^{(theo,N)} = 2\sqrt{\frac{E\gamma}{Nt}} \qquad (5)$$

Taking $E = 1\text{TPa}$ (Young's modulus of graphene), $\gamma = 0.2\,\text{N/m}$ (surface energy of graphene; however note that in reality $\gamma$ could be also larger as a consequence of additional dissipative mechanisms, e.g. fracture and friction in addition to adhesion), the predicted maximum strength for single walled nanotubes ($N=1$) is:

$$\sigma_C^{(\max)} = \sigma_C^{(theo,1)} = 48.5\text{GPa} \qquad (6)$$

whereas for double or triple walled nanotubes $\sigma_C^{(theo,2)} = 34.3\text{GPa}$ or $\sigma_C^{(theo,3)} = 28.0\text{GPa}$.

Considering the previous calculations in *Report 1* for the equilibrium contact length *a* during polygonization, we can write:

$$\sigma_C = 2\cos\beta\sqrt{E\gamma\frac{6a + P_{vdW}}{S}} \tag{7}$$

Furthermore, the ratio between the bundle strengths with ($\sigma_C^{(0)}$) or without ($\sigma_C^{(O)}$) self-collapsed nanotubes, is predicted to be:

$$\frac{\sigma_C^{(0)}}{\sigma_C^{(O)}} = \sqrt{\frac{2\pi R + P_{vdW}}{2\pi R\left(1 - \frac{1}{R}\sqrt{\frac{N^\alpha D}{2\gamma}}\right) + P_{vdW}}}, \quad \text{for } R \geq R_C^{(N)} = \sqrt{\frac{3N^\alpha D}{\gamma}} \tag{8}$$

The maximal strength increment induced by the self-collapse is thus:

$$\left.\frac{\sigma_C^{(0)}}{\sigma_C^{(O)}}\right|_{max} = \sqrt{\frac{1}{1 - \frac{1}{\sqrt{6}}}} \approx 1.30 \tag{9}$$

Eq. (9) shows that the self-collapse could enhance the nanotube bundle strength up to ~30%.

*Toughness*

The energy dissipated during the fracture of the bundle is:

$$W_C \approx F_C \ell \tag{10}$$

where $\ell$ is the mean nanotube length and (before separation a sliding of length $\sim \ell$ occurs at a constant force $\sim F_C$). Accordingly, the effective fracture energy per unit area of the bundle is:

$$G_C \approx \sigma_C \ell \tag{11}$$

Taking $\sigma_C = 10\text{GPa}$ and $\ell = 1\mu\text{m}$ gives $G_C \approx 10^4 \text{ N/m}$, corresponding to a facture toughness of:

$$K_C = \sqrt{G_C E} \tag{12}$$

of the order of $K_C \approx 100\text{MPa}\sqrt{\text{m}}$ ($E = 1\text{TPa}$). The energy per unit area is high but not proportional to the bundle length, thus suggesting a quasi-brittle, rather than ductile, behaviour. Eq. (11) suggests to increase the nanotube length, in order to increase the toughness; in the limit of coincident nanotube and bundle lengths (still assuming the sliding failure mode, practically a composite bundle would be more appropriate in order to diffuse the damage in the entire bundle volume prior to fracture), the dissipated energy, per unit volume or mass, becomes:

$$J_{CV}^{(\max)} \approx \sigma_C, \ J_{CM}^{(\max)} \approx \frac{\sigma_C}{\rho} \tag{13}$$

where $\rho$ is the material density, reached at a failure strain of $\varepsilon_C \approx 100\%$. Taking $\sigma_C = 10\text{GPa}$ and $\rho \approx 1000\,\text{Kg}/\text{m}^3$ yields a maximum dissipated energy per unit mass of $\sim 10\text{MJ}/\text{g}$, enormously higher than that of spider silk ($\sim 165\text{J}/\text{g}$). Even if the predictions of eqs. (13) are fully ideal, they suggest that we are far from such a limit for toughness and thus that super-tough composites can be produced in the near future (much more easily than super-strong composites).

*Hierarchical bundle*

Experimentally, three hierarchies of structure within the fibre can be observed (Motta et al., Advanced Materials, 2007, 19, 3721): doubly walled nanotubes with diameters in the range 5-10nm, bundles of 20-60nm diameter composed by self-collapsed nanotubes and, finally, the macroscopic fibre composed by bundles with preferred orientation along the fibre axis, see Figure 1.

In such a case eq. (4) is still valid, but the proper value of the ratio $\frac{P}{S}$ must be evaluated, according to the hierarchical nature of the fibre. Consider two sub-bundles, having surface energy $\gamma$, Young's modulus $E$, Poisson's ratio $v$ and radius $R$; the JKR theory of adhesion gives the contact width as (see [7]):

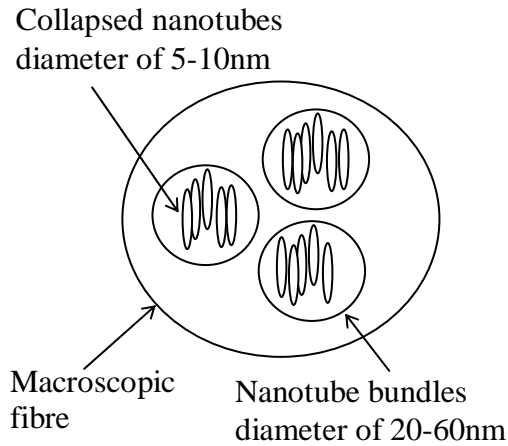

Figure 1: Scheme of the cross-section of the hierarchical nanotube fibre.

$$w = 4\left(\frac{4R^2\gamma(1-v^2)}{\pi E}\right)^{\frac{1}{3}} \tag{14}$$

Assuming an hexagonal packing of the bundles within the fibre and independent contact widths, we deduce:

$$\frac{P}{S} = \frac{24\left(\frac{4R^2\gamma(1-v^2)}{\pi E}\right)^{\frac{1}{3}}}{\pi R^2} \tag{15}$$

Finally, the strength for the bundle is predicted according to:

$$\sigma_C = 2\cos\beta\sqrt{E\gamma\frac{P}{S}} = 2\cos\beta\sqrt{E\gamma\frac{24\left(\frac{4R^2\gamma(1-v^2)}{\pi E}\right)^{\frac{1}{3}}}{\pi R^2}} = \\ = \frac{2\sqrt{24}(4(1-v^2))^{1/6}}{\pi^{2/3}}\cos\beta E^{1/3}\gamma^{2/3}R^{-2/3} \approx 5.76\cos\beta E^{1/3}\gamma^{2/3}R^{-2/3} \tag{16}$$

Taking $E = 1\text{TPa}$, $v = \beta = 0$, $\gamma = 0.2\,\text{N/m}$, $R = 10-30\text{nm}$, we deduce $\frac{S}{P} \approx 112\text{nm}$ and thus $\sigma_C^{(theo)} \approx 2.04 - 4.24\text{GPa}$. These values are in close agreement to the experimental observations performed by the Windle's group, even if we have taken for $E, v, \beta, \gamma$ just plausible values rather than independently measured or fitting values.

The analysis suggests to reduce the sub-bundles radius to increase the strength, as well as to increase the nanotube length to enlarge the toughness, but only up to $\ell_C$ in order not to be detrimental for the strength.

## 3. The role of the solvent on the polygonization, collapse, self-collapse and "dog-bone" configurations of an isolated nanotube or of nanotubes in a bundle

*Polygonization*

The solvent around a bundle of radius $R_B$ induces an additional pressure $\frac{\gamma_t}{R_B}$, according to the Laplace's equation, where $\gamma_t$ is the solvent surface tension. Accordingly, the equation of state for the polygonization becomes:

$$p(a^*) = \frac{3N^\alpha D}{ca^*\left(1-\frac{3}{\pi}a^*\right)^2 R^3} - \frac{\gamma_t}{R_B} - \frac{6\gamma_{wet}}{ca^*R} \tag{1}$$

where $\gamma_{wet}$ is the surface energy of the nanotubes in the presence of the solvent.

Without solvent and external pressure ($\gamma_t = p = 0$, $\gamma_{wet} \to \gamma_{dry}$) the equilibrium is reached in the following configuration:

$$a^*_{0dry} = a^*(p = \gamma_t = 0) = \frac{\pi}{3}\left(1 - \frac{1}{R}\sqrt{\frac{N^\alpha D}{2\gamma_{dry}}}\right) \tag{2}$$

In the presence of solvent and without an additional external pressure, the new equilibrium contact length can be calculated according to eq. (1) posing $p = 0$ and solving the corresponding third-order polynomial equation. However, a correction with respect to $a^*_{0dry}$ can be deduced inserting into eq. (1) $a^* = a^*_{0dry}(\gamma_{dry} \to \gamma_{wet})(1 + \varepsilon_a)$, neglecting the powers of $\varepsilon_a \ll 1$ higher than one and recalling eq. (2); we accordingly find:

$$\varepsilon_a = \frac{\pi c \gamma_t}{36 R_B \gamma_{wet}} \sqrt{\frac{N^\alpha D}{2\gamma_{wet}}} \tag{3}$$

Thus, the equilibrium configuration in the presence of solvent is:

$$a^*_{0wet} = a^*(p=0) \approx \frac{\pi}{3}\left(1 + \varepsilon_a - \frac{1}{R}\sqrt{\frac{N^\alpha D}{2\gamma_{wet}}}\right) \tag{4}$$

and since $\varepsilon_a \ll 1$, the only significant effect of the presence of the solvent is $\gamma_{dry} \to \gamma_{wet}$, namely $a^*_{0wet} \approx a^*_{0dry}(\gamma_{dry} \to \gamma_{wet})$.

*Collapse*

The critical pressure in the presence of solvent becomes:

$$p_C = \frac{3N^\alpha D}{R^3} - \frac{\gamma_{wet}}{R} - \frac{\gamma_t}{R_B} \tag{5}$$

*Self-collapse*

From eq. (5) we derive the following condition for the self-collapse, i.e. collapse under zero pressure, of a nanotube in a bundle in the absence of solvent ($\gamma_t = p = 0$, $\gamma_{wet} \to \gamma_{dry}$):

$$R \geq R^{(N)}_{Cdry} = \sqrt{\frac{2N^\alpha D}{\gamma_{dry}}} \tag{6}$$

In the presence of solvent and without an additional external pressure, the new self-collapse radius can be calculated according to eq. (5) posing $p = 0$ and solving the corresponding third-order polynomial equation. However, a correction with respect to $R^{(N)}_{Cdry}$ can be deduced inserting into eq. (5) $R = R^{(N)}_{Cdry}(\gamma_{dry} \to \gamma_{wet})(1 + \varepsilon_R)$, neglecting the powers of $\varepsilon_R \ll 1$ higher than one and recalling eq. (6); we accordingly find:

$$\varepsilon_R = -\frac{\gamma_t}{2 R_B \gamma_{wet}} \sqrt{\frac{3 N^\alpha D}{\gamma_{wet}}} \tag{7}$$

Thus, the self-collapse radius in the presence of solvent is:

$$R \geq R_{Cwet}^{(N)} = \sqrt{\frac{3N^\alpha D}{\gamma_{wet}}}(1+\varepsilon_R) \qquad (8)$$

and since $\varepsilon_R \ll 1$, the only significant effect of the presence of the solvent is $\gamma_{dry} \to \gamma_{wet}$, namely $R_{Cwet}^{(N)} \approx R_{Cdry}^{(N)}(\gamma_{dry} \to \gamma_{wet})$.

In contrast, for an isolated nanotube the critical pressure becomes:

$$p_C = \frac{3N^\alpha D}{R^3} - \frac{\gamma_t}{R} \qquad (9)$$

and thus the role of the solvent is crucial; the self-collapse takes place for:

$$R \geq R_C^{(N)} = \sqrt{\frac{3N^\alpha D}{\gamma_t}} \qquad (10)$$

Taking $D = 0.11\,\text{nN} \cdot \text{nm}$ (bending stiffness of graphite), $\gamma_t = 0.07\,\text{N/m}$ (surface tension of water) we find $2R_C^{(1)} \approx 4.3\,\text{nm}$. Considering an intermediate coupling between the walls ($\alpha \approx 2$), the critical diameters for double and triple walled nanotubes are $2R_C^{(2)} \approx 8.7\,\text{nm}$ and $2R_C^{(3)} \approx 13.0\,\text{nm}$. Note that for solvent with low surface tension the critical diameter could become quite large, for example for argon $\gamma_t = 0.005 - 0.015\,\text{N/m}$ and $2R_C^{(1)} \approx 9.4 - 16.2\,\text{nm}$.

*"Dog-bone" configuration*

In the presence of solvent the state equation for the dog-bone configuration becomes:

$$p(r) = \frac{N^\alpha D}{2r^3} - \frac{\gamma}{r} - \frac{\gamma_t}{R_B} \qquad (11)$$

where $\gamma = \gamma_{in} + \gamma_{out}$ is the total, sum of the inner and outer, surface energy; if the nanotube is not in contact with other surrounding nanotubes $\gamma_{out} = 0$; moreover $\gamma_{in,out} = \gamma_{dry,wet}$ as a consequence of the presence and/or absence inside and/or outside the nanotube of the solvent.

For zero surface energy and tension the equilibrium pressure is positive (inward), whereas complimentary for zero bending stiffness it is negative (outward). Under zero pressure and without solvent ($p = \gamma_t = 0$) the equilibrium is reached for:

$$r_{0dry} = \sqrt{\frac{N^\alpha D}{2\gamma}} \qquad (12)$$

In the presence of solvent and without an additional external pressure, the new equilibrium radius can be calculated according to eq. (11) posing $p = 0$ and solving the corresponding third-order polynomial equation. However, a correction with respect to the previously evaluated case of absence of solvent can be deduced inserting into eq. (11) $r = r_{0dry}(1+\varepsilon_r)$, neglecting the powers of $\varepsilon_r \ll 1$ higher than one and recalling eq. (12); we accordingly find:

$$\varepsilon_r = -\frac{1}{4R_B}\sqrt{\frac{2N^\alpha D}{\gamma}} \qquad (13)$$

Thus the equilibrium configuration in the presence of solvent is:

$$r_{0wet} = \sqrt{\frac{N^\alpha D}{2\gamma}}(1+\varepsilon_r) \qquad (14)$$

and since $\varepsilon_r \ll 1$, the only significant effect of the presence of the solvent is the modification of $\gamma$ (in which $\gamma_{dry} \to \gamma_{wet}$).

Posing the limiting condition of $r=R/2$ in eq. (11), we deduce the critical pressure corresponding to the dog-bone "opening" (note that the process is stable, even if rapid around a particular configuration, see Report 1):

$$p_O = \frac{4N^\alpha D}{R^3} - \frac{2\gamma}{R} - \frac{\gamma_t}{R_B} \qquad (15)$$

Moreover, posing $p_O = 0$ in eq. (15) in absence of solvent ($p_O = \gamma_t = 0$) suggests that for nanotube radii:

$$R \le \sqrt{\frac{2N^\alpha D}{\gamma}} = 2r_{0dry} \qquad (16)$$

the "dog-bone" configuration cannot be self-maintained.

In the presence of solvent the new critical radius can be calculated according to eq. (15) posing $p_O = 0$ and solving the corresponding third-order polynomial equation. However, a correction with respect to the previously evaluated case of absence of solvent can be deduced inserting into eq. (16) $R = 2r_{0dry}(1+\varepsilon)$, neglecting the powers of $\varepsilon \ll 1$ higher than one and recalling eq. (16); we accordingly find:

$$\varepsilon = -\frac{\gamma_t}{2R_B\gamma}\sqrt{\frac{N^\alpha D}{2\gamma}} \qquad (17)$$

Thus the equilibrium configuration in the presence of solvent is:

$$R \le \sqrt{\frac{2N^\alpha D}{\gamma}}(1+\varepsilon) \qquad (18)$$

and since $\varepsilon \ll 1$, the only significant effect of the presence of the solvent is the modification of $\gamma$ (in which $\gamma_{dry} \to \gamma_{wet}$).

In contrast, for an isolated nanotube, $R_B \to r$ (the solvent pressure acts on the two lobes of radius $r$) and $\gamma_{out} = 0$, thus:

$$p(r) = \frac{N^\alpha D}{2r^3} - \frac{\gamma_{in} + \gamma_t}{r} \tag{19}$$

Accordingly, all the derived previous equations remain valid with formally $\gamma = \gamma_{in} + \gamma_t$ and $R_B \to \infty$. Accordingly, the equilibrium is reached for:

$$r_0 = \sqrt{\frac{N^\alpha D}{2\gamma}} \tag{20}$$

and the dog-bone configuration cannot be self-maintained for:

$$R \leq 2r_0 \tag{21}$$

## 4. Collapse pressure and self-collapse of peapods, fullerite crystals and fullerenes

*Peapods*

In the case of peapods, the collapse pressure is increased as a consequence of the presence inside the nanotube of the fullerenes; since the critical pressure of fullerenes is much higher than that of a nanotube (see in the following), we treat the peapod as a nanotube of finite length $L$, equal to the (centre-centre) distance between two adjacent fullerenes. Note that the classical buckling formula of cylindrical shells assumes infinite length.
According to elasticity (see for instance R. M. Jones, Buckling of bars, plates and shells, Bull Ridge Publishing, Blacksburg, Virginia, USA, 2006) for a long cylinder the buckling pressure is:

$$p_c = \frac{3N^\alpha D}{R^3}, \quad L \gg L_c \tag{1}$$

whereas for short cylinders:

$$p_c = \frac{4\pi^2 N^\alpha D}{RL^2}, \quad L \ll L_c \tag{2}$$

(Note that the proportionality observed by the Elliot's group of $p_C \propto R^{-1}$ seems to suggest a short nanotube behaviour, even if we must recall that the computational results were obtained asymptotically in the limit of large $L$, which suggests this coincidence as incidental).
The critical length governing the transition can be calculated equating eqs. (2) and (3):

$$L_c = \frac{2\pi}{\sqrt{3}} R \tag{3}$$

For intermediate lengths, elasticity poses:

$$p_c = \frac{\pi^2 \sqrt{\sqrt{1-v^2}} N^\alpha D}{RL\sqrt{Rt}}, \quad L \sim L_c \tag{4}$$

Comparing eqs. (1) or (2) with eq. (4) one would deduce the critical lengths governing the transition, from the short or long to the intermediate lengths.

Moreover, we expect the following pressure generated by the surrounding bundle and/or solvent:

$$p_\gamma \approx \frac{\gamma}{R} + \frac{\gamma_t}{R_B} \tag{5}$$

The presence of the solvent around the bundle does not affect significantly the critical pressure, since $R_B \gg R$, whereas for isolated peapod $R_B = R$; accordingly:

$$p_\gamma \approx \frac{\gamma + \gamma_t}{R} \tag{6}$$

is valid for both peapods in a bundle (with $\gamma_t = 0$ and $\gamma$ effective surface energy, thus "wet" in the presence of solvent or "dry" if absent) as well as for an isolated peapod (with $\gamma = 0$). Revisiting the previous elastic results, we thus expect for the buckling of peapods the following regimes:

$$p_C = \frac{3N^\alpha D}{R^3} - \frac{\gamma + \gamma_t}{R}, \quad L \gg L_c \tag{7a}$$

$$p_C = \frac{\pi^2 \sqrt{\sqrt{1-v^2}} N^\alpha D}{RL\sqrt{Rt}} - \frac{\gamma + \gamma_t}{R}, \quad L \sim L_c \tag{7b}$$

$$p_C = \frac{4\pi^2 N^\alpha D}{RL^2} - \frac{\gamma + \gamma_t}{R}, \quad L \ll L_c \tag{7c}$$

Let us introduce the fullerene content as:

$$f = \frac{2R}{L} \tag{8}$$

the previous equation become:

$$p_C = \frac{3N^\alpha D}{R^3} - \frac{\gamma + \gamma_t}{R}, \quad f \ll f_c = \frac{\sqrt{3}}{\pi} \tag{9a}$$

$$p_C = \frac{\pi^2 \sqrt{\sqrt{1-v^2}} N^\alpha D}{2R^2 \sqrt{Rt}} f - \frac{\gamma + \gamma_t}{R}, \quad f \sim f_c \tag{9b}$$

$$p_C = \frac{\pi^2 N^\alpha D}{R^3} f^2 - \frac{\gamma + \gamma_t}{R}, \quad f \gg f_c \tag{9c}$$

This behaviour is summarized in Figure 1.

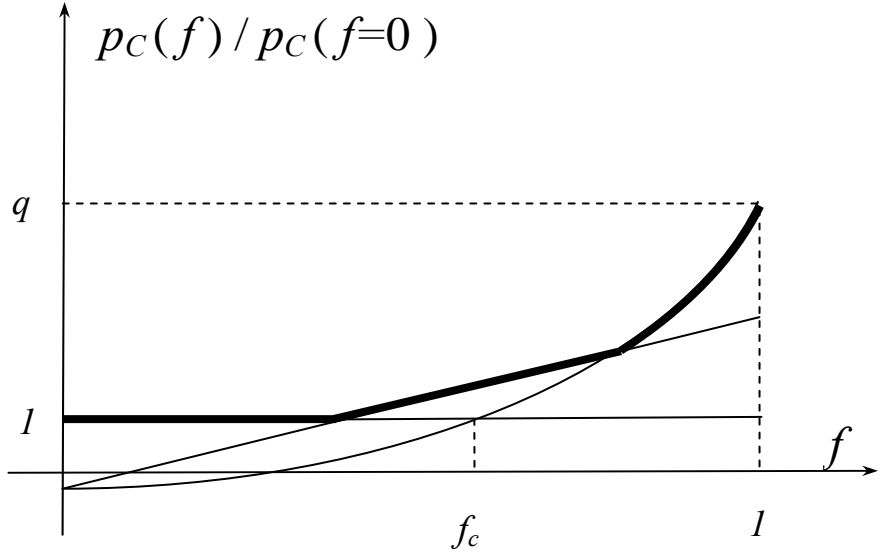

Figure 1: Theoretical dependence of the buckling pressure versus fullerene content.

We can estimate the ratio $q$ between the buckling pressures for $f=0$ and $f=1$, as:

$$q = \frac{p_C(f=1)}{p_C(f=0)} = \frac{\pi^2 - \frac{(\gamma+\gamma_t)R^2}{N^\alpha D}}{3 - \frac{(\gamma+\gamma_t)R^2}{N^\alpha D}} \qquad (10)$$

Noting that in the treated case $\frac{(\gamma+\gamma_t)R^2}{N^\alpha D} \ll 1$, we expect $q \approx \pi^2/3$, as numerically observed (private communication by Elliot).

From eq. (7) we derive the following conditions for the self-collapse, i.e. collapse under zero pressure:

$$R_C^{(N)} = \sqrt{3}\sqrt{N^\alpha}\sqrt{\frac{D}{\gamma+\gamma_t}}, \quad L \gg L_c \qquad (11a)$$

$$R_C^{(N)} L_C^{(N)2} = \frac{\pi^4 D^2 \sqrt{1-v^2}}{(\gamma+\gamma_t)^2 t}, \quad L \sim L_c \qquad (11b)$$

$$L_C^{(N)} = \sqrt{2\pi}\sqrt{N^\alpha}\sqrt{\frac{D}{\gamma+\gamma_t}}, \quad L \ll L_c \qquad (11a)$$

Note that for small fullerene content the self-collapse is dictated by a critical radius, as for empty nanotubes, whereas for large fullerene content the self-collapse is dictated by a critical distance between two adjacent fullerenes (in the intermediate case length and radius are comparable).

*Fullerite crystals and fullerenes*

Similarly, the critical pressure of fullerenes in a fullerite crystal is:

$$p_C = \frac{2}{\sqrt{3(1-v^2)}} \frac{N^\alpha E t^2}{R^2} - \frac{2\gamma}{R} - \frac{2\gamma_t}{R_B} \quad (12)$$

where $1 \leq \alpha \leq 2$ described the interaction between the walls; the first term is that posed by elasticity (that considers $\alpha = 2$; see for instance A.V. Pogorevol, *Bending of Surface and Stability of Shells*, American Mathematical Society, Providence, Rhode Island, USA, 1988), whereas the factor of two in the surface energy and tension is expected according to the Laplace's equation.

The presence of the solvent around the crystal does not affect significantly the critical pressure, since $R_B \gg R$, whereas for isolated fullerene $R_B = R$; accordingly:

$$p_C = \frac{2}{\sqrt{3(1-v^2)}} \frac{N^\alpha E t^2}{R^2} - \frac{2(\gamma + \gamma_t)}{R} \quad (13)$$

is valid for both fullerenes in fullerite (with $\gamma_t = 0$ and $\gamma$ effective surface energy, thus "wet" in the presence of solvent or "dry" if absent) as well as for an isolated fullerene (with $\gamma = 0$).

If the fullerenes are larger than the nanotubes, the interaction pressure (for the calculation see N. Pugno, *Thermomechanical stresses in fullerenes@nanotube*. J. OF NANOMATERIALS, 2008, 156724, 5 pp.) could cause their collapse. Note that the factor $(t/R)^2$ (for fullerenes), appearing in stead of $(t/R)^3$ (for nanotubes), shows that the critical pressure for fullerenes is much higher than that for nanotubes, at least for $t/R \ll 1$.

From eq. (13) we derive the following condition for the self-collapse, i.e. collapse under zero pressure of fullerenes in crystals or isolated:

$$R_C^{(N)} = \frac{\lambda N^\alpha E t^2}{2(\gamma + \gamma_t)} \quad (14)$$

Note that for $v = 0$, $N = 1$, $E = 1\text{TPa}$, $t = 0.34\text{nm}$, $2(\gamma + \gamma_t) = 0.4\,\text{N/m}$, we find $R_C^{(1)} \approx 334\text{nm}$, showing that fullerenes are highly stable and thus that peapods with high fullerene concentrations are ideal solution against nanotube buckling (the result of eq. (4) is similar to that observed during the adhesion of nanovectors on a substrate, see N. Pugno, *Flexible nanovectors*. J. OF PHYSICS – COND. MAT., 2008, 20, 474205, 7pp).

**Conclusions**

We have discovered that the influence of the surrounding nanotubes in a bundle is similar to that of a liquid having surface tension equal to the surface energy of the nanotubes. This surprising behaviour is confirmed by the calculation of the self-collapse diameters of nanotubes in a bundle. Other systems, such as peapods, fullerites, are similarly treated, including the effect of the presence of a solvent. Finally, we have evaluated the strength and toughness of the nanotube bundle, with or without collapsed nanotubes, assuming a sliding failure.